\documentclass[aps,prb,superscriptaddress,preprint,showpacs]{revtex4}  

\usepackage{amsmath}
\usepackage{graphicx}

\newcommand{\beq}{\begin{equation}} 
\newcommand{\eeq}{\end{equation}}

\newcommand{\beqa}{\begin{eqnarray}} 
\newcommand{\eeqa}{\end{eqnarray}}

\begin{document} 

\title{LCHO-CI method for the voltage control of exchange interaction
  in gated lateral quantum dot networks} 

\author{Irene Puerto Gimenez} 
\affiliation{Quantum Theory Group, 
Institute for Microstructural Sciences, 
National Research Council, Ottawa, Canada K1A 0R6}
\affiliation{ 
Department of Fundamental Physics, 
University of La Laguna, 
Tenerife, Spain}

\author{Marek Korkusinski} 
\affiliation{Quantum Theory Group, 
Institute for Microstructural Sciences, 
National Research Council, Ottawa, Canada K1A 0R6} 

\author{Pawel Hawrylak} 
\affiliation{Quantum Theory Group, 
Institute for Microstructural Sciences, 
National Research Council, Ottawa, Canada K1A 0R6} 

\begin{abstract} 
We present a computational LCHO-CI approach allowing for the 
simulation of exchange interactions in  gated lateral  quantum 
dot networks. 
The approach is based on single-particle states calculated using a 
linear combination of harmonic orbitals (LCHO) of each of the dots,  
and a configuration interaction (CI) approach to the interacting 
electron problem.   
The LCHO-CI method is applied to a network of three quantum dots with one 
electron spin per dot, and a Heisenberg spin Hamiltonian  is 
derived. 
The manipulation of spin states of a three-spin molecule by applying 
bias to one of the dots is demonstrated and related to the bias
dependence of effective exchange interaction parameters. 
\end{abstract} 


\maketitle

\section{Introduction} 

The electron spin is a quantum two level system, a natural candidate  
for a qubit.\cite{nielsen_chuang_book,brum_hawrylak_sm1997,loss_divincenzo_pra1998} 
For this reason there is currently significant interest in coupling of 
individual electron spins localized at different spatial locations to 
realize quantum 
gates.\cite{brum_hawrylak_sm1997,loss_divincenzo_pra1998,divincenzo_bacon_nature2000} 
The coupling of the spins of localized interacting electrons is 
investigated using the Heisenberg 
Hamiltonian.\cite{divincenzo_bacon_nature2000}   
In this approach, the Coulomb interactions between electrons are 
reduced to the exchange coupling of their spins. 
The exchange interaction is parametrized by the exchange constants $J$ 
dependent upon the specific implementation of the system. 

In lateral gated quantum dot devices the spatial localization of individual 
electrons can be achieved by electrostatic coupling to their charge.  
To date, controlled confinement of one, two, and three spatially 
separated electrons has been demonstrated using the 
single,\cite{ciorga_sachrajda_prb2000,tarucha_austing_prl1996}  
double,\cite{holleitner_blick_science2002,pioro_abolfath_prb2005,koppens_folk_science2005,petta_johnson_science2005,hatano_stopa_science2005} 
and triple\cite{gaudreau_studenikin_prl2006,korkusinski_gimenez_prb2007}   
quantum dots, respectively. 
In these devices the electronic orbital degrees of freedom are 
manipulated directly by tuning the gate voltages, which, only through 
topology and statistics, translates into the control over the total 
spin of the  
system. 
\cite{mizel_lidar_prl2004,mizel_lidar_prb2004,woodworth_mizel_jpcm2006,hawrylak_korkusinski_ssc2005,weinstein_hellberg_pra2005,scarola_park_prl2004} 
The voltage control of exchange coupling of two electrons in a single 
dot \cite{kyriakidis_pioro_prb2002} and  
coherent control of spin states of a two-electron double-dot molecule
\cite{petta_johnson_science2005} was recently demonstrated.  
The Heisenberg Hamiltonian appropriate for this two-spin system 
is parametrized by a single exchange constant $J$, whose magnitude 
defines the energy gap between the spin singlet and triplet eigenstates. 
By measuring this gap one can establish the dependence of $J$ 
on the gate voltages. In a more complex network of $N$ quantum dots
with one spin per dot the number of exchange couplings needed is equal
to the number $N ( N-1)/2$ of pairs.  
Even the simplest network of  quantum dots, a  triple quantum dot 
with one electron per dot, 
is described by a Heisenberg Hamiltonian with  three exchange 
constants, depending nontrivially on the geometry of the system and
the gate layout. 
This dependence was studied, e.g., by Scarola and Das 
Sarma,\cite{scarola_dassarma_pra2005} who used the Hubbard,   
variational, and exact diagonalization approaches to demonstrate that  
the three-spin model is valid only for a limited range of triple-dot 
parameters.   
Mizel and  
Lidar\cite{mizel_lidar_prl2004,mizel_lidar_prb2004,woodworth_mizel_jpcm2006}  
arrived at similar conclusions using the Heitler-London and  
Hund-M\"ulliken schemes to calculate the energy levels  
of three coupled dots with one electron per dot.  
In both cases the many-body effects were responsible for the  
appearance of higher-order terms in the effective spin Hamiltonian.  
The above studies were performed on triple-dot systems consisting of 
three dots on resonance, and did not account for tuning of individual 
dots.  In Ref.~\onlinecite{korkusinski_gimenez_prb2007} 
we used the Hubbard model, and in 
Ref.~\onlinecite{hawrylak_korkusinski_ssc2005} - the real-space wave 
functions coupled with the configuration-interaction technique (RSP-CI) to 
analyze the voltage-tunable three-electron gated lateral triple-dot 
device, however without mapping the resulting electronic properties 
onto the three-spin model. 
The Hubbard model is simple but requires parametrization 
of the Hamiltonian and cannot be directly linked to gates and applied
voltages.  
The RSP-CI technique is very accurate but is difficult to implement
for the future simulation of the time evolution of the quantum system.

The purpose of this work is to present a different computational
approach, where linear combination of harmonic orbitals (LCHO) of each
of the dots is used to describe tunneling of electrons, and the
configuration interaction (CI) approach is used for the treatment of
exchange and correlation in the interacting electron problem.  
The LCHO-CI method can then be directly used for the derivation of the
effective Heisenberg Hamiltonian consistent with gate voltages, and
for the simulation of the quantum operations.   

We illustrate our method by analyzing the tunability of the exchange 
interaction and manipulation of the three spin system with voltage in 
a triangular quantum dot molecule.    
The triangular quantum dot molecule, realized recently by Gaudreau et 
al., \cite{gaudreau_studenikin_prl2006} is needed both for 
implementation of the quantum teleportation and creation of
three-particle maximally entangled GHZ state.   

\section{LCHO-CI method for the electronic structure of quantum dot
  networks}  

\subsection{Single electron states in a quantum dot network} 

We describe here the LCHO method for the calculation of single
electron states in a network of  two-dimensional quantum dots. 
The method is illustrated on the example of a triple quantum dot. 

\begin{figure}[h]  
\includegraphics[width=0.8\textwidth]{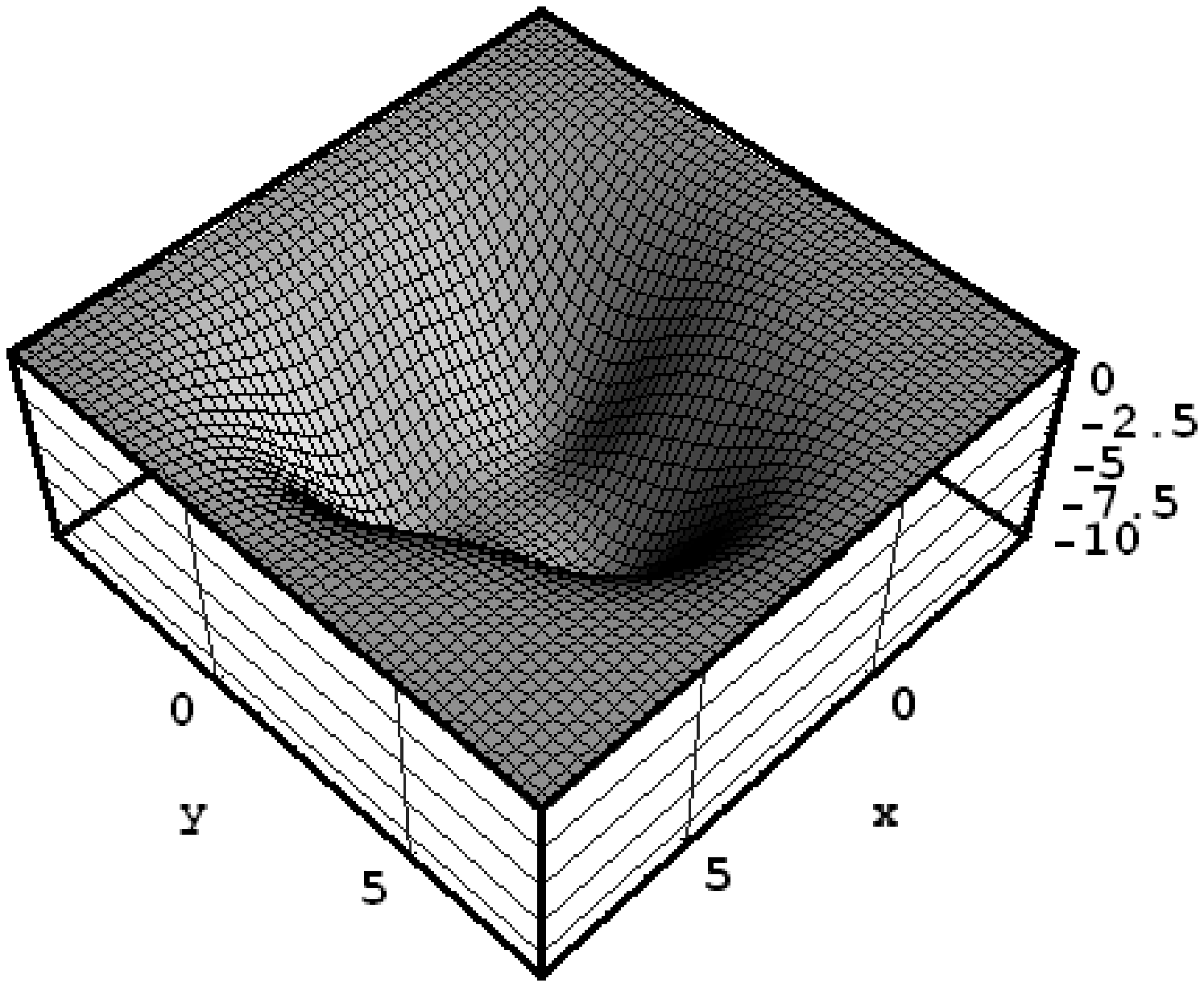}
\caption{3D plot of the triple quantum dot potential with equal dots
  characterized by depths $V_i = 10 \ Ry^*$  and widths ${d}_{i} = 2.3
  \ a_B^*$.  
  The dots are centered at (0,0), (4,0) and (2,3.4641)  
  forming an equilateral triangle with side lengths $4 \ a_B^*$.} 
\label{fig:3QDplot} 
\end{figure}

We consider a triple quantum dot molecule created electrostatically by 
lateral gates\cite{gaudreau_studenikin_prl2006} and approximate its
lateral confinement potential by a sum of three two-dimensional
Gaussians,  
$ 
V^{3QD} = - \sum_{i=1}^{3} V_{i}  
\exp \left[ 
{-\frac   {({x}-{x}_{i})^2 + ({y}-{y}_{i})^2}  {{d}_{i}^2 }   }     
\right] , 
$ 
with ${x}_{i}$ and ${y}_{i}$ being the coordinates of the center of
$i$th Gaussian ($i$=1,2,3), ${V}_{i}$ being its depth, and ${d}_{i}$
being its characteristic width. 
The depth and width of each Gaussian are functions of the gate
voltages.  
The centers (${x}_{i}$, ${y}_{i}$) of each dot are arbitrary but in 
the rest of the paper they are chosen to lie in the 
corners of an equilateral triangle. 
In the following we express all distances in effective Bohr radii 
$a_B^*= \varepsilon \hbar^2 / m^* e^2$ 
and all energies in units of effective Rydberg 
$Ry^*= e^2/2 \varepsilon a_0^*$, 
where $e$ and $m^*$ are the electron charge and effective mass, 
respectively, and $\varepsilon$ is the dielectric constant of the material. 
GaAs parameters $m^*=0.067$ $m_0$ and $\varepsilon=12.4$ give 
$Ry^* = 5.93$ meV and $a_B^* = 97.9$ $\AA$. 
Figure~\ref{fig:3QDplot} shows the triple-dot confining potential for 
$V_1=V_2=V_3=10$ $Ry^*$ and $d_1=d_2=d_3=2.3$ $a_B^*$. 
The dimensionless Hamiltonian for one electron in the potential 
of the triple quantum dot network is written in the following form:
\begin{equation}\label{eq:1e} 
H =  
-\frac{\partial^2}{\partial x^2} - \frac{\partial^2}{\partial y^2} 
- \sum_{i=1}^{3} V_{i} 
\exp \left[ {-\frac{({x}-{x}_{i})^2 + ({y}-{y}_{i})^2}  {{d}_{i}^2}}
\right].
\end{equation} 
Since an exact analytical solution of the eigenvalue problem of the 
Hamiltonian (\ref{eq:1e}) is not known, an approximate method is 
needed to calculate the energies and eigenvectors of one electron in 
the quantum dot network.   
Here we employ a quantum dot analog of the  Linear Combination of
Atomic Orbitals (LCAO) method.  
Expanding the Gaussian potential of dot $i$ to second order in $\vec{r}$, 
$ 
-V_i \exp \left[ -{\frac{(\vec{r}-\vec{r}_i)^2}{d_i^2}} 
\right] \approx -V_i + V_i \frac{(\vec{r}-\vec{r}_i)^2}{d_i^2}, 
$ 
results in the harmonic oscillator (HO) potential  
$ 
-V_i + \frac{1}{4} \Omega_i^2 (\vec{r}-\vec{r}_i)^2 
$, with 
$\Omega_i = 2\sqrt{V_i} /d_i$. 
The eigenfunctions of this HO potential, 
$ 
\phi^i_{n m}(x,y)= \varphi^i_{n}(x-x_i) \varphi^i_{m}(y-y_i), 
$ 
are products of the 1D HO eigenfunctions 
$ 
\varphi_{n}(t)=   \left(  \frac{1}{\pi l^2}               \right) ^ \frac{1}{4}  
                  \left(  \frac{l^{2n}}{2^{n} n!}         \right) ^ \frac{1}{2}                 
                  \left(  \frac{t}{l^2}-\frac{d}{dt}      \right) ^ n  
           \exp   \left(  -\frac{t^2}{2 l^2}              \right), 
$ 
where $l=l_i=\sqrt{\frac{2}{\Omega_i}}=\left( \frac{d_i^2}{V_i} 
\right) ^\frac{1}{4}$. 
The first three states have the explicit form 
$ 
\varphi_{0}(t)=  \left(  \frac{1}{\pi l^2}  \right) ^ \frac{1}{4}   
 \exp  \left(  -\frac{t^2}{2 l^2}  \right), 
$ 
$ 
\varphi_{1}(t)=  \sqrt{2} \left(  \frac{1}{\pi l^2}       \right) ^ \frac{1}{4}  
                          \left(  \frac{t}{l}             \right)    
           \exp   \left(  -\frac{t^2}{2 l^2}              \right) , 
$ 
$ 
\varphi_{2}(t)=  \frac{1}{2\sqrt{2}} \left(  \frac{1}{\pi l^2}  \right) ^ \frac{1}{4}  
                          \left( -2 + \frac{4t^2}{l^2}    \right)             
           \exp   \left(  -\frac{t^2}{2 l^2}              \right) . 
$

Hence, the molecular eigenfunctions of Eq.~(\ref{eq:1e}), $\vert \xi_n
\rangle$, are written as linear combinations of the HO orbitals (LCHO)
centered on each dot:  
\begin{equation} \label{eq:ev} 
\vert \xi_n \rangle = \sum_{i=1}^{3 n_{o}} a_i^n \vert \phi_i \rangle ,  
\end{equation} 
where $a_i^n$ are the expansion coefficients and $n_{o}$ is the number of  
HO orbitals per dot. 
To simplify notation, the three indices $i$, $n$, and $m$ of the  
HO eigenfunctions $\phi^i_{nm}(x,y)$ have been replaced by 
the composite index $i$. 
Substituting this expression into the eigenvalue problem of Eq.~(\ref{eq:1e})
and multiplying on the left by $\langle \phi_j \vert$ gives 
\begin{equation}  \nonumber 
 \sum_{i=1}^{3 n_{o}} \langle \phi_j \vert H \vert \phi_i \rangle a_i^n 
= \epsilon_{n} \sum_{i=1}^{3 n_{o}}\langle \phi_j \vert \phi_i \rangle  a_i^n. 
\end{equation} 
Defining $H_\phi$ as the Hamiltonian matrix in LCHO basis with
elements 
$ \langle \phi^j_{00} | H |  \phi^i_{00} \rangle  = 
\int{d \vec{r} \phi^{j*}_{00}(x,y) H \phi^i_{00}(x,y) } $ 
and $S_\phi$ as the overlap matrix with elements
$\langle \phi_{00}^j \vert \phi_{00}^i \rangle = 
\int{d \vec{r} \phi^{j*}_{00}(x,y) \phi^i_{00}(x,y) } $ 
allows us to write the generalized eigenvalue problem as 
\begin{equation} \label{eq:GenEig} 
H_{\phi} \vec{a}^n = \epsilon_{n} S_{\phi} \vec{a}^n . 
\end{equation} 
Now, defining a new vector  
\begin{equation} \label{eq:ab} 
\vec{b}^n = (\sqrt{S_\phi}) \vec{a}^n    
\end{equation} 
and multiplying Eq.~(\ref{eq:GenEig}) by $(\sqrt{S_\phi})^{-1}$ on the left   
gives the standard eigenvalue problem 
\begin{equation}  \label{eq:eigeq} 
(\sqrt{S_\phi})^{ -1} H_\phi (\sqrt{S_\phi})^{ -1} \vec{b}^n =
\epsilon_n \vec{b}^n .  
\end{equation} 
In order to calculate $(\sqrt{S_\phi})^{-1}$ the eigenvalue problem of
the overlap matrix   
$ 
S_\phi V_S = V_S E_S 
$ 
is solved. 
Here $V_S$ is the matrix with the eigenvectors and $E_S$ is the
diagonal matrix with the eigenvalues. 
Once the values of these two matrices are calculated, 
$(\sqrt{S_\phi})^{-1}$  is obtained from 
$ 
(\sqrt{S_\phi})^{-1} =V_S E_S^{-1/2} V_S^T . 
$ 
Then the energies $\epsilon_n$ (\ref{eq:GenEig})  and eigenvectors
(\ref{eq:ev}) of the electron in the quantum dot network can be
calculated using Eqs.~(\ref{eq:eigeq}) and (\ref{eq:ab}).  
The accuracy of the solution  depends on the number $3n_{o}$ of HO
orbitals included in the basis.   
Increasing the number of basis states increases the accuracy
of results.   

We now analyze the matrix elements of the Hamiltonian $H_\phi$ in the
basis $
\{|\phi_{00}^1\rangle,|\phi_{00}^2\rangle,|\phi_{00}^3\rangle\}$
showing explicitly the various contributions to the diagonal onsite
energies  and off-diagonal tunneling matrix elements: 
\beq 
H_\phi= 
\left[ 
  \begin{array}{ccc}     
        \epsilon^d_1 & t_{12} & t_{13} \\ 
        t_{21} & \epsilon^d_2 & t_{23} \\ 
        t_{31} & t_{32} & \epsilon^d_3 \\ 
  \end{array} 
\right] . 
\eeq 
If we label each of the three dots with indices $i, j, k$, we can
express the diagonal matrix elements as  
\begin{eqnarray}
\epsilon^d_i & = & 
\langle \phi^i_{00} | H | \phi^i_{00} \rangle \nonumber \\ 
& = & 
(-V_i + \Omega_i ) + \langle \phi^i_{00} | \delta V^i | \phi^i_{00}
\rangle  
+ \langle \phi^i_{00} | V^j | \phi^i_{00} \rangle +
\langle \phi^i_{00} | V^k | \phi^i_{00} \rangle , 
\label{eq:epsilond}
\end{eqnarray}
where we denote 
$- V_{i}\exp \left[ {-\frac{(x-x_{i})^2+(y-y_{i})^2}{d_{i}^2}}\right]
= V^i_{HO} + \delta V^i $, 
with $V^i_{HO}$ being the 2D harmonic oscillator potential associated
with dot $i$:  
$V^i_{HO} = -V_i + \frac{1}{4} \Omega_i^2 (\vec{r}-\vec{r}_i)^2 $. 
The first term in (\ref{eq:epsilond}), $(-V_i + \Omega_i )$, is the
dominant term.  
It is the energy of the ground state of the harmonic oscillator
potential $V^i_{HO}$.  
The second term, $\langle \phi^i_{00} | \delta V^i | \phi^i_{00}
\rangle$, gives the correction due to non-harmonicity of the confining
potential.  
The third and fourth terms,  
$\langle \phi^i_{00} | V^j | \phi^i_{00} \rangle$ 
and $\langle \phi^i_{00} | V^k | \phi^i_{00} \rangle$,  
give the correction to the energy level of an isolated quantum dot $i$
due to the presence of the other two quantum dot potentials. 

The off-diagonal matrix elements describe electron tunneling from dot
$i$ to dot $j$.  
The tunneling matrix elements are determined by several contributions : 
\begin{eqnarray}  
t_{ji} & =&  \langle \phi^j_{00} | H | \phi^i_{00} \rangle\nonumber\\
& = & (-V_i + \Omega_i ) \langle \phi^j_{00} | \phi^i_{00} \rangle +
\langle \phi^j_{00} | \delta V^i | \phi^i_{00} \rangle
+ \langle \phi^j_{00} | V^j | \phi^i_{00} \rangle + 
\langle \phi^j_{00} | V^k | \phi^i_{00} \rangle . 
 \label{eq:t}
\end{eqnarray}
The first term is directly proportional to the overlap between HO
wave functions centered on different dots,  
$ \langle \phi^j_{00} | \phi^i_{00} \rangle $, and is therefore small
for dots which are far apart.  
The second term,  $\langle \phi^j_{00} | \delta V^i |
\phi^i_{00}\rangle$,  
is the correction due to non-harmonicity of the confining potential. 
The third term, $\langle \phi^j_{00} | V^j | \phi^i_{00} \rangle $,
 is a ``two-centered'' integral, as it involves a product of three
 functions centered at two different dots.  
The fourth  term, $ \langle \phi^j_{00}| V^k  | \phi^i_{00} \rangle$, 
is a ``three-centered'' integral being thus the smallest term. 
 This analysis of $H_\phi$ matrix elements can also be applied to a
larger HO  basis including more shells ($p$, $d$, $\dots$).   


\subsection{Many-electron states in a quantum dot network} 

The many-electron Hamiltonian of the quantum dot network  written in
second quantization is
\begin{equation} \label{eq:hamiltNe2q} 
\hat{H} = \sum_{j} \epsilon_{j} c_{j}^{+} c_{j} 
+\frac{1}{2} \sum_{ijkl}  \langle ij \lvert v \rvert kl \rangle
c_{i}^{+} c_{j}^{+} c_{k} c_{l},  
\end{equation} 
where $c^+_m$ and $c_m$ are, respectively,  the creation and
annihilation operators of a particle on the spin-orbital itinerant
(molecular) state $m$,    
$\psi_m(\vec{r}) = \xi_{n}(\vec{r}) \chi_{m_s}$, 
where $\chi_{m_s}$ is the electronic spinor corresponding to the spin 
$m_s=\pm1/2$. 
Indices $i,j,k$ and $l$ run over spin-orbitals, 
$\epsilon_{j}$ is the single-particle energy of an electron in the 
molecular state $j$   
and $v$ is the dimensionless Coulomb potential, $\frac{2}{\lvert 
  \vec{r}_{2}-\vec{r}_{1} \rvert}$.   

The Coulomb matrix elements (CMEs) in the itinerant basis 
are computed as linear combinations of CMEs in the localized HO basis  
\begin{equation} \label{eq:CME} 
\langle ij \lvert v \rvert kl \rangle  =  
\langle \chi_i |  \chi_l \rangle \langle  \chi_j | \chi_k \rangle  
\sum_{r=1}^{3 n_{o}}  \sum_{s=1}^{3 n_{o}}  
\sum_{t=1}^{3 n_{o}}  \sum_{u=1}^{3 n_{o}} 
a_r^i  a_s^j  a_t^k  a_u^l \langle r s \lvert v \rvert t u \rangle, 
\end{equation} 
where 
$ 
\langle r s \lvert v \rvert t u \rangle = \int d \vec{r}_1   
\int d \vec{r}_2 
\phi_r^*(\vec{r}_1) \phi_s^*(\vec{r}_2)
\frac{2}{\lvert \vec{r}_{2}-\vec{r}_{1} \rvert} 
\phi_t(\vec{r}_2)\phi_u(\vec{r}_1). 
$ 
Using the identity 
$\frac{2}{| \vec{r}_1 - \vec{r}_2 |} =  
\frac{1}{\pi}   \int {d \vec{q} \over q }
\exp{ \left[ i \vec{q}( \vec{r}_1 - \vec{r}_2 ) \right] }, 
$ we have 
\begin{equation}   \label{eq:CMEHO} 
 \langle r s \lvert v \rvert t u \rangle  
 = \frac{1}{\pi}   \int_{0}^{\infty}{dq} \int_{0}^{2 \pi}{d \vartheta } 
 F_{ru}^+ (q,\vartheta)  G_{ru}^+ (q,\vartheta)  
F_{st}^- (q,\vartheta) G_{st}^- (q,\vartheta) 
\end{equation} 
with 
$ 
F_{nm} ^{\pm} (q,\vartheta) = \int_{-\infty}^{\infty} dt \varphi^*_{n}(t) 
\varphi_{m}(t)  \exp{ \left[ \pm i t q \cos{\vartheta}  \right] } 
$ 
and 
$ 
G_{nm} ^{\pm} (q,\vartheta) = \int_{-\infty}^{\infty} dt \varphi^*_{n}(t) 
\varphi_{m}(t)  \exp{ \left[ \pm i t q \sin{\vartheta}  \right] } 
$. 
These four integrals were obtained analytically and 
the double integral (\ref{eq:CMEHO}) over $q$ and $\vartheta$ was 
carried out numerically. 

In order to obtain the energy levels and coherent eigenfunctions of
$N$ electrons in the quantum dot network we combine the LCHO method
with the configuration interaction (CI) approach.  
In the CI method we build the $N$-electron basis  out of all possible
configurations of the $N$ electrons on the molecular single-particle
states.   
The many-electron  eigenvalues and eigenfunctions  are obtained by
diagonalizing the Hamiltonian  (\ref{eq:hamiltNe2q}) in this
$N$-electron basis of configurations.  
The number of these configurations $n_C$  (and hence the dimension of
$H$) depends on the number of electrons $N$ and the number $n_{SO}$ of
molecular spin-orbital states through   
$n_C = \frac{n_{SO}!}{(n_{SO}-N)!N!} $ . 
In the case of as few as $N=3$ electrons, LCHO-CI calculations using
only $s$-type  HO orbital per dot gives $6$ spin-orbitals, leading to
$n_C=20$. 
If for LCHO calculation we also include $p$-type HO orbitals, there  
are $18$ molecular spin-orbitals, giving  $n_C=816$, while 
including also $d$ HO orbitals gives $n_C=7140$.  
The number of configurations in the 3-electron basis increases very 
rapidly with increasing number of HO orbitals in the one-electron 
calculations thus increasing the computational requirements. 

One is primarily interested in quantum networks with one electron (one
spin) per quantum dot. 
If one retains only one orbital per dot, the number of spin-orbitals
is $n_{SO}=2 N$, and number of possible configurations is reduced to  
$ 
n_C = \frac{(2N)!}{N!N!} . 
$ 
For $N=10$ spins we already have $184756$ configurations. 

Further reduction in the number of configurations is possible. 
Since the Hamiltonian (\ref{eq:hamiltNe2q}) is rotationally invariant,
the Hamiltonian needs to be diagonalized only in one of the subspaces
of lowest $|S_z|$.    
In what follows  we present numerical results for the three-electron 
$S_z=-1/2$ and $S_z=-3/2$ states obtained with the LCHO-CI method.

\section{Effective Heisenberg model} 

In order to establish a connection with quantum computation, it is 
convenient to approximate the Hamiltonian (\ref{eq:hamiltNe2q}) of the
system of singly-occupied dots in a quantum dot network by an
effective Heisenberg spin Hamiltonian.   
For three electrons in three quantum dots this Hamiltonian takes the
form   
\begin{equation} \label{eq:Heisenberg} 
H^{s} =  {J}_{12} \frac{1}{4} \vec{\sigma}_1 \vec{\sigma}_2 +  
{J}_{23} \frac{1}{4} \vec{\sigma}_2 \vec{\sigma}_3 +  
{J}_{13} \frac{1}{4} \vec{\sigma}_1 \vec{\sigma}_3, 
\end{equation} 
where ${J}_{ij}$ are the dimensionless exchange coupling constants,
and $\vec{\sigma}$  are the Pauli matrices.    
From this Hamiltonian the quantum gate $\sqrt{SWAP}$ can be obtained by 
turning on one $J_{ij}$ for an appropriate amount of
time.\cite{loss_divincenzo_pra1998}
Combining this gate with single qubit gates we can obtain the $CNOT$
gate, and $CNOT$ together with one-qubit gates form the universal
basis for quantum computing.\cite{nielsen_chuang_book}

Since $H^{s}$ commutes with $S_z$, $H^{s}$ matrix is 
block diagonal in the basis of eigenvectors of $S_z$. 
Here we will treat the subspace $S_z=-1/2$. 
Taking the basis vectors  
$\{ 
| \downarrow \downarrow \uparrow \rangle , 
| \downarrow \uparrow \downarrow \rangle , 
| \uparrow \downarrow  \downarrow  \rangle \}
$ 
 we obtain 
\begin{equation} \label{eq:1/2} 
H^{s} = {1\over4}\left[ 
    \begin{array}{ccc} 
      J_{12}-J_{23}-J_{13} & 2J_{23} & 2J_{13}\\ 
      2J_{23}  &  -J_{12}-J_{23}+J_{13} & 2J_{12} \\ 
      2J_{13}  &  2J_{12}  &  -J_{12}+J_{23}-J_{13} \\ 
    \end{array} 
    \right]. 
\end{equation} 

Let us introduce a Jacobi basis of spin states: 
$|\beta_a \rangle = \frac{1}{\sqrt{2}}(|\downarrow \downarrow \uparrow \rangle 
- |\downarrow \uparrow \downarrow \rangle  )$, 
$|\beta_b \rangle = \frac{1}{\sqrt{6}}(|\downarrow \downarrow \uparrow\rangle 
+ |\downarrow \uparrow \downarrow \rangle - 2 |\uparrow  \downarrow \downarrow \rangle )$, 
and 
$|\beta_c \rangle = \frac{1}{\sqrt{3}}(|\downarrow \downarrow \uparrow \rangle 
+ |\downarrow \uparrow \downarrow \rangle + 
| \uparrow \downarrow \downarrow \rangle )$ . 
The Heisenberg Hamiltonian (\ref{eq:Heisenberg}) in the Jacobi basis
has the following form:
\begin{equation} \label{eq:1/2-2} 
H^{s} = 
\left[ 
\begin{array}{ccc} 
  -\frac{3}{4} J_{av} -\frac{3}{4} ( J_{23}-J_{av} ) & \frac{\sqrt{3}}{4} (J_{12}-J_{13}) & 0    \\ 
  \frac{\sqrt{3}}{4} (J_{12}-J_{13})   &  -\frac{3}{4} J_{av} + \frac{3}{4} ( J_{23}-J_{av} ) & 0   \\ 
  0  &  0    &   \frac{3}{4} J_{av}  \\ 
    \end{array} 
\right], 
\end{equation} 
where $J_{av}=(J_{12}+J_{23}+J_{13})/3 $ is the average exchange constant. 
We see that if all exchange constants are equal the two states
$|\beta_a\rangle$ and $|\beta_b\rangle$ are degenerate eigenstates 
with energy $ -\frac{3}{4} J_{av} $ and total spin $S=1/2$,
while the state $| \beta_c\rangle$, corresponding to total spin
$S=3/2$, has energy $+\frac{3}{4} J_{av} $.  
The Hamiltonian of the two degenerate states  $|\beta_a\rangle$ and
$|\beta_b\rangle $ is analogous to the  Hamiltonian of a single spin
in a magnetic field, where the magnetic field in the $z$ direction is
proportional to $\frac{3}{4} ( J_{23}-J_{av} ) $ and the magnetic
field in the  $x$ direction is proportional to $\frac{\sqrt{3}}{4}
(J_{12}-J_{13})$.
This is why our system of three electrons in three dots can be thought
of as a single coded qubit, whose logical states can be manipulated by
varying the exchange constants.\cite{hawrylak_korkusinski_ssc2005}

\section{Results for a triple quantum dot network} 

We illustrate our theoretical approach on an example of a triple dot molecule with parameters given in Fig. \ref{fig:3QDplot}.  
In section \ref{1el} we present the energies and eigenfunctions of one
electron in the triple dot molecule calculated with LCHO method. 
In section \ref{3el} the energies of three electrons in the triple dot
molecule obtained with LCHO-CI method are analyzed and compared with
results obtained with Hubbard model.\cite{korkusinski_gimenez_prb2007}
In section \ref{bias} we analyze the triple dot molecule
with one and three electrons when one of the dots is being biased.  
Finally, in section \ref{heis} we use the Heisenberg model to study
the effects of biasing of one dot in a triple dot molecule with three
electrons and compare the results with those obtained using LCHO-CI
method.

\subsection{One electron in a triple quantum dot \label{1el}} 

\begin{figure}[h]
\includegraphics[width=0.8\textwidth]{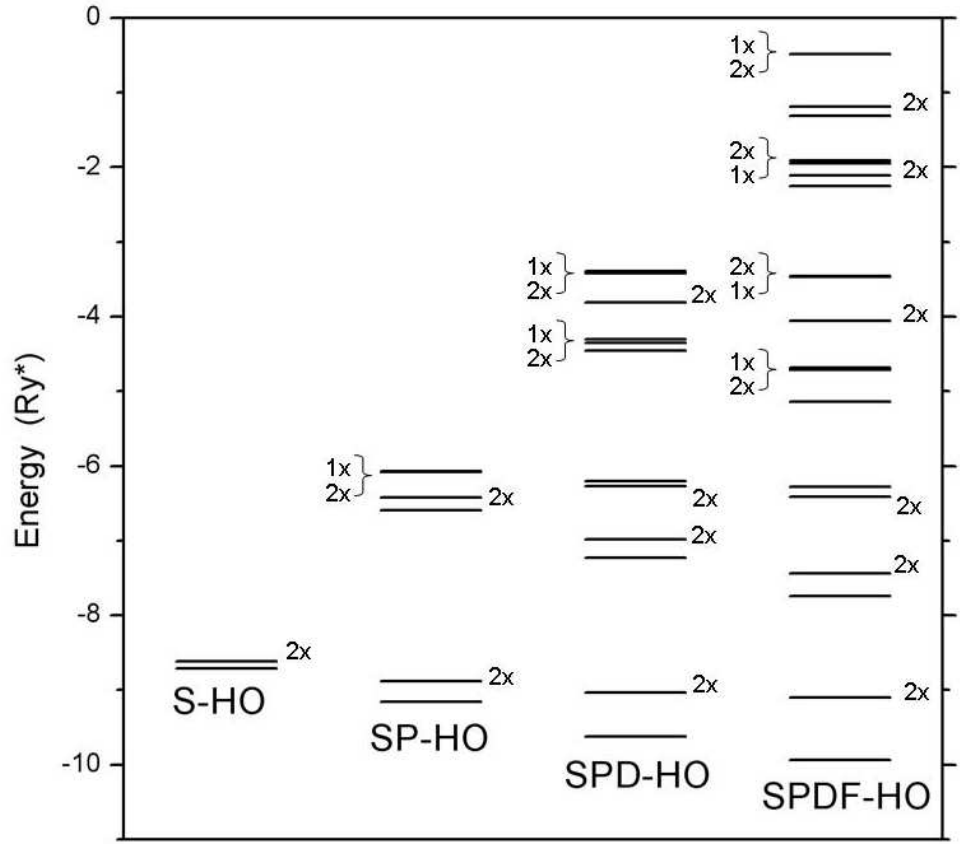}
\caption{Energy levels of one electron in the triple quantum dot
  potential of Fig. \ref{fig:3QDplot} calculated with the LCHO method
  as a function of  the number of HO basis states .}  
\label{fig:1elspectrum3QD} 
\end{figure}

Figure \ref{fig:1elspectrum3QD} shows the set  of one-electron energy
levels in the triple quantum dot potential of Fig.~\ref{fig:3QDplot}  
obtained by solving Eq.~(\ref{eq:GenEig}) for different LCHO basis sets: 
$s: \{ |\phi_{00}^1\rangle,|\phi_{00}^2\rangle,|\phi_{00}^3\rangle 
\}$, 
$s+p:\{ |\phi_{00}^1\rangle,  |\phi_{10}^1\rangle,  |\phi_{01}^1\rangle,  
|\phi_{00}^2\rangle,  |\phi_{10}^2\rangle,  |\phi_{01}^2\rangle, 
|\phi_{00}^3\rangle,  |\phi_{10}^3\rangle,  |\phi_{01}^3\rangle  \}$, 
$s+p+d$, and $s+p+d+f$. 
The number of orbitals per shell in the spectra of a triple quantum
dot  is three times the number of orbitals per shell in the spectra of
one quantum dot.  
Thus, we have three $s$ orbitals, six $p$ orbitals, 
nine $d$ orbitals and so on.  

\begin{figure}[h]
\includegraphics[width=0.8\textwidth]{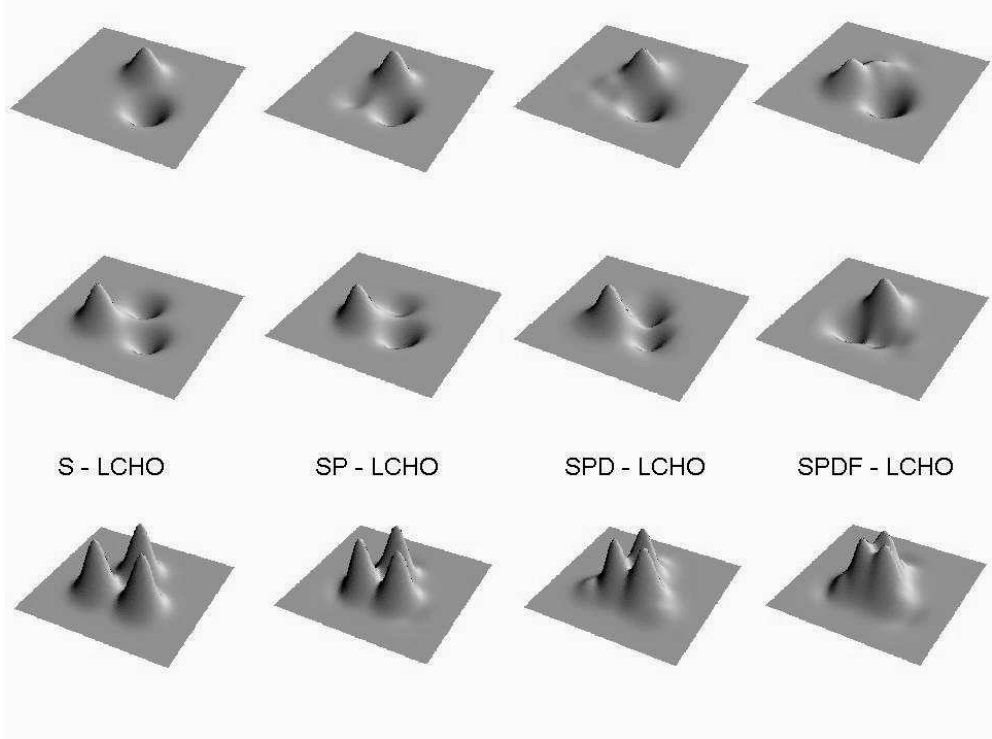}
\caption{Comparison of the three lowest energy eigenfunctions ($\vert
  \xi_1 \rangle$, $\vert \xi_2 \rangle$ and $\vert \xi_3 \rangle$)  
  of one electron in a triple quantum dot, calculated with the 
  LCHO method using different numbers of HO wave functions as basis.  
  Eigenfunctions at the bottom correspond to the ground state, $\vert
  \xi_1 \rangle$, and the two at the top, $\vert \xi_2 \rangle$ and
  $\vert \xi_3 \rangle$, correspond to the degenerate excited states.}  
\label{fig:1elwf} 
\end{figure}

The spectrum of one electron in the HO potential associated with one of
the dots has electronic shells  
with level spacing $\Omega_i = 2\sqrt{V_i}/d_i = 2.75$~$Ry^*$. 
For the electron in an isolated Gaussian dot and assuming LCHO
basis sets composed of $sp$, $spd$ and $spdf$ HO shells one obtains $s$-$p$
level spacings of $2.44$, $2.46$ and $2.37$ $Ry^*$ respectively.
When the electron moves in the potential of the triple dot, the
tunneling hybridizes shells from different dots and leads to molecular
levels.  
Comparing the spectra of one electron in the triple-dot molecule for 
increasing number of HO shells   
in the LCHO basis we find that as the number of shells increases the 
low energy molecular levels decrease their energy and converge to a
definite value.    
The ground state of one electron in the symmetric triple-dot molecule is  
nondegenerate, and the first excited state is doubly degenerate.  
Figure \ref{fig:1elwf} shows the eigenfunctions of these three levels 
($s$ orbitals $\xi_1 (\vec{r}), \xi_2 (\vec{r})$ and $\xi_3(\vec{r})$) 
calculated using HO basis sets with different number of orbitals. 
Results obtained from calculations with only $s$-type HO orbitals in
the LCHO basis are  qualitatively the same as those obtained when more
HO  orbitals ($p$, $d$, $f$) are included.   
Based on this conclusion, and anticipating a larger number of
electrons in a quantum dot network, we will build the configurations
for the three electron  problem including one-electron energies and
eigenfunctions obtained from calculations with the $s$-LCHO basis. 
Here we show typical numerical values for this $s$-LCHO calculation. 
The off-diagonal overlap matrix elements appearing in
Eq.~(\ref{eq:GenEig}) are $s = \langle \phi^j_{00} | \phi^i_{00}
\rangle  = 0.004088 $.  
The  diagonal on-site energies defined in Eq.~(\ref{eq:epsilond}) are
$\epsilon^d=-8.647493$~$Ry^*$  
and have the contributions ($ i \neq j \neq k$): 
$ (-V_i + \Omega_i ) = -7.250193 $~$Ry^*$, 
$ \langle \phi^i_{00} | \delta V^i | \phi^i_{00} \rangle  = -0.166187$~$Ry^*$, 
$ \langle \phi^i_{00} | V^j | \phi^i_{00} \rangle  =
-0.615556$~$Ry^*$, and  
$ \langle \phi^i_{00} | V^k | \phi^i_{00} \rangle  =  -0.615556$~$Ry^*$. 
On the other hand, the tunneling matrix elements hybridizing the
atomic orbitals of adjacent dots, defined in Eq.~(\ref{eq:t}),
are $t=-0.067162$~$Ry^*$ and are composed of ($ i \neq j \neq k$): 
$(-V_i + \Omega_i ) \langle \phi^j_{00} | \phi^i_{00} \rangle  =
-0.029641$~$Ry^*$,  
$ \langle \phi^j_{00} | \delta V^i | \phi^i_{00} \rangle  =
-0.014140$~$Ry^*$,  
$ \langle \phi^j_{00} | V^j | \phi^i_{00} \rangle  = -0.018489$~$Ry^*$ and 
$ \langle \phi^j_{00} | V^k | \phi^i_{00} \rangle  = -0.004892$~$Ry^*$. 
After solving Eq.~(\ref{eq:GenEig}), three  energy levels
corresponding to hybridized $s$ shell and their eigenfunctions are
obtained.  
The $s$-shell energy gap obtained  between the ground state, $\vert \xi_1
\rangle$, and the degenerate excited state,  
$\vert \xi_2 \rangle$ and $\vert \xi_3 \rangle$, 
 is $\Delta \epsilon_S = 0.095$~$Ry^*$.
We plot these energies in Fig.~\ref{fig:1elspectrum3QD} (first
spectrum on the left).  
The eigenfunctions of these three levels, plotted in the left-hand
part of Fig.~\ref{fig:1elwf}, are:  
\begin{eqnarray} 
\vert \xi_1 \rangle &=& 1/\sqrt{3(1+2 s)}(\vert \phi_1 \rangle + \vert
\phi_2 \rangle + \vert \phi_3 \rangle ), \nonumber \\  
\vert \xi_2 \rangle &=& 1/\sqrt{2(1-s)} (\vert \phi_1 \rangle  - \vert
\phi_2 \rangle  ), 
 \label{eq:1elef} \\  
\vert \xi_3 \rangle &=& 1/\sqrt{6(1-s)} (\vert \phi_1 \rangle +\vert
\phi_2 \rangle - 2\vert \phi_3 \rangle ),\nonumber
\end{eqnarray} 
where $s$ is the off-diagonal overlap matrix element. 
This number appears because the basis eigenfunctions are not orthogonal.

\subsection{Three electrons in a triple quantum dot network \label{3el}}

\begin{figure}[h]  
\includegraphics[width=0.8\textwidth]{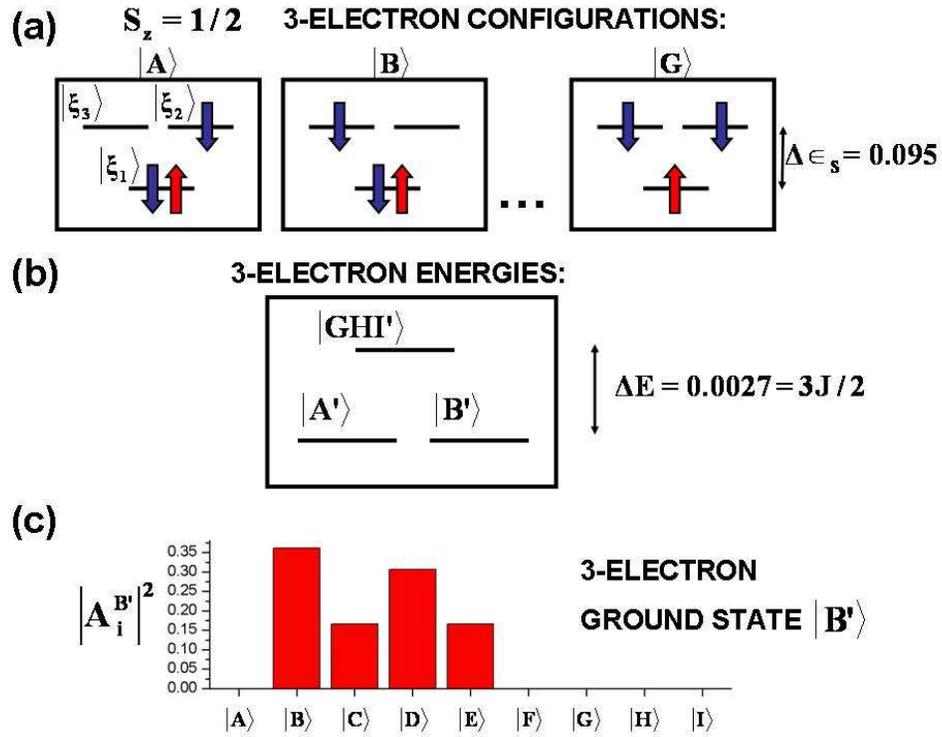}
\caption{(a) Examples of three-electron configurations that form the
  basis for CI calculations.  
  (b) The lowest group of levels of the three electron spectrum. 
  (c) Contribution of different configurations to one of the two
  states of the ground level.}  
\label{fig:3elanalysis} 
\end{figure}

Once the one-electron problem has been solved, we proceed to solve the
three-electron Hamiltonian (\ref{eq:hamiltNe2q}) using the configuration
interaction method (CI).  
The first step is to build the basis of three-electron configurations
in which $H$ is to be diagonalized. 
This is done for the subspaces $S_z=-1/2$ and $S_z=-3/2$. 
There are nine three-electron configurations with $S_z=-1/2$ obtained
by distributing the three electrons among the six  molecular
spin-orbitals $\{ |\xi_{i} \rangle | \downarrow (\uparrow ) \rangle,
i=1,2,3 \}$ (for illustration see Fig. \ref{fig:3elanalysis} (a)).
These configurations can be grouped into six doubly-occupied
configurations:  
$\vert A \rangle = c^+_{1\downarrow} c^+_{2\downarrow} c^+_{1\uparrow}
\vert 0 \rangle$,   
$\vert B \rangle = c^+_{1\downarrow}c^+_{3\downarrow}c^+_{1\uparrow}
\vert 0 \rangle$,  
$\vert C \rangle = c^+_{1\downarrow}  c^+_{2\downarrow}
c^+_{2\uparrow} \vert 0 \rangle$,  
$\vert D \rangle = c^+_{2\downarrow}  c^+_{3\downarrow}
c^+_{2\uparrow} \vert 0 \rangle$,
$\vert E \rangle = c^+_{1\downarrow}  c^+_{3\downarrow}
c^+_{3\uparrow} \vert 0 \rangle$, and  
$\vert F \rangle = c^+_{2\downarrow}  c^+_{3\downarrow}
c^+_{3\uparrow} \vert 0 \rangle$, 
and three singly-occupied configurations: 
$\vert G \rangle = c^+_{2\downarrow}  c^+_{3\downarrow}
c^+_{1\uparrow} \vert 0 \rangle$,
$\vert H \rangle = c^+_{1\downarrow}  c^+_{3\downarrow}
c^+_{2\uparrow} \vert 0 \rangle$, and
$\vert I \rangle = c^+_{1\downarrow}  c^+_{2\downarrow}
c^+_{3\uparrow} \vert 0 \rangle$.  
For the spin-polarized system with $S_z=-3/2$ there is only one
possible configuration,
$ \vert K \rangle = c^+_{1\downarrow}  c^+_{2\downarrow}
c^+_{3\downarrow} \vert 0 \rangle$.

\begin{figure}[h]  
\includegraphics[width=0.8\textwidth]{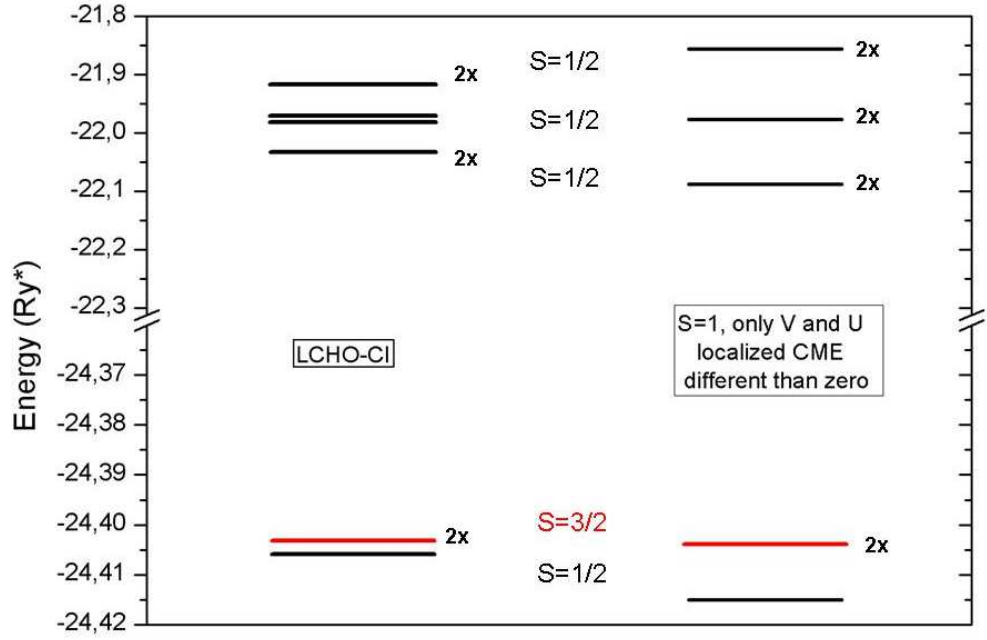}
\caption{Left: spectrum of three electrons with $S_z =-1/2$ in the
  triple quantum dot calculated with LCHO-CI method. 
  Right: spectrum of three electrons with $S_z =-1/2$ in the triple dot
  calculated with LCHO-CI method but neglecting the overlap matrix and
  all CMEs except for $U=\langle rr | v | rr \rangle $ and $V=\langle
  rs | v | sr \rangle $. }   
\label{fig:Hubbard} 
\end{figure}

The construction of the Hamiltonian matrix (\ref{eq:hamiltNe2q}) in
these bases of configurations requires the knowledge of Coulomb matrix
elements.  
The Coulomb matrix elements in the itinerant basis are computed from
Coulomb matrix elements in the localized basis as defined in
Eq.~(\ref{eq:CME}), using the coefficients of the one-electron molecular
eigenfunctions written in HO basis. 
The localized CMEs, Eq.~(\ref{eq:CMEHO}),  for the triple quantum dot
potential of Fig. \ref{fig:3QDplot} consist of the onsite repulsion 
$U=\langle rr | v | rr \rangle = 2.939179$~$Ry^*$, 
interdot repulsion $V=\langle rs | v | sr \rangle = 0.512857$~$Ry^*$,
and a number of small electron-electron scattering terms: 
$\langle rr | v | rs \rangle = 0.004653$~$Ry^*$, 
$\langle rs | v | st \rangle = 0.003446$~$Ry^*$, 
$\langle rs | v | rs \rangle = 0.000049$~$Ry^*$, 
and $\langle rr | v | st \rangle = 0.000019$~$Ry^*$. 
Upon rotation into the itinerant basis we find that the largest
Coulomb matrix elements include direct repulsion on the lowest kinetic
energy level  
$\langle 11 | v | 11 \rangle = 1.315811$~$Ry^*$
and on the degenerate excited state $\langle 22 | v | 22 \rangle =
\langle 33 | v | 33 \rangle = 1.730884$~$Ry^*$.  
The direct and exchange interaction terms between the ground and
excited states are $\langle 12 | v | 21 \rangle = 1.320137$~$Ry^*$ and
$\langle 12 | v | 12 \rangle = 0.806983$~$Ry^*$, respectively,
while the scattering term
$\langle 11 | v | 22 \rangle = 0.806983$~$Ry^*$. 
Similar matrix elements are obtained for 
the degenerate shell of excited states: 
$\langle 23 | v | 32 \rangle = 0.918383$~$Ry^*$, 
$\langle 23 | v | 23 \rangle = 0.406251$~$Ry^*$, and 
$\langle 22 | v | 33 \rangle = 0.406251$~$Ry^*$. 
These Coulomb matrix elements are used in the construction of the
three-electron Hamiltonian (\ref{eq:hamiltNe2q}). 
After diagonalizing this Hamiltonian in the basis 
of configurations with $S_z=-1/2$: $\{|i\rangle \}= \{ | A \rangle, | B
\rangle, | C \rangle, | D \rangle, | E \rangle, | F \rangle, | G \rangle, | H
\rangle, | I \rangle \}$,  
we obtain a spectrum of nine levels, shown in Fig.~\ref{fig:Hubbard}
(left). 
The nine eigenfunctions $| i' \rangle$ are linear combinations of the
basis configurations $ | i' \rangle = \sum_{i=A}^I  A_i^{i'} | i
\rangle $.   
In the spectrum we can clearly distinguish two groups of levels
separated by a large gap.  
The group in the upper part of the spectrum is composed of six energy
levels and the group with lowest energy has three levels. 
We focus on this low-energy group. 
It is composed of a doubly degenerate ground state and a
non-degenerate first excited state, as shown in
Fig. \ref{fig:3elanalysis} (b), separated by the energy gap 
$\Delta E = 0.0027$~$Ry^*$.
The doubly degenerate ground state has total spin $S=1/2$ and  
its two eigenfunctions $| A' \rangle$ and $| B' \rangle$ have their
biggest contribution from the three electron configurations $\vert A
\rangle$ and $\vert B \rangle$,  respectively.  
However, these states are highly correlated, with large contributions
from other configurations.  
The eigenfunction $| A' \rangle$, with biggest contribution from the 
state $\vert A \rangle$, has also contributions from 
configurations $\vert F \rangle$,$\vert H \rangle$ and $\vert I \rangle$:  
$| A' \rangle = 0.6014 \vert A \rangle + 0.5533 \vert F \rangle
-0.4069 \vert H \rangle -0.4082 \vert I \rangle$,  
while the eigenfunction $| B' \rangle$ with biggest contribution from
state $\vert B \rangle$,  has also contributions from states $\vert C
\rangle$, $\vert D \rangle$ and $\vert E \rangle$:  
$ | B' \rangle = 0.6014 \vert B \rangle -0.4075 \vert C \rangle
-0.5533 \vert D \rangle +0.4075 \vert E \rangle $, as shown in
Fig.~\ref{fig:3elanalysis} (c).   
The first excited state of this $S_z=-1/2$ three-electron spectrum,
state $| GHI' \rangle$, has total $S=3/2$.  
Its eigenfunction $| GHI' \rangle$ has approximately equal
contribution from the three singly occupied molecular configurations:
$\vert G \rangle$, $\vert H \rangle$ and $\vert I \rangle$.  
Taking into account all four $S_z$ subspaces, the ground state is four
times  degenerate (two levels with $S_z=-1/2$ and two levels with
$S_z=1/2$) and the first excited state is also four times degenerate
(one level belonging to each $S_z$ subspace).

In order to compare the microscopic model with the Hubbard
model\cite{korkusinski_gimenez_prb2007} we propose a modified LCHO-CI
method.
First, we carry out the $s$-LCHO one-electron calculations neglecting
the overlap matrix.  
The resulting one-electron spectrum also shows a non-degenerate ground
state and a doubly degenerate first excited state but with a bigger
energy gap $\Delta \epsilon_S^{Hub} = 3 t = 0.201$~$Ry^*$.  
Next, we apply the CI method to the $S_z=-1/2$ subspace of Hamiltonian
(\ref{eq:hamiltNe2q}) but neglect all localized CME terms different
than $U$ and $V$. 
As in the itinerant electron basis, the three-electron spectrum
obtained in this calculation shows two groups of levels, shown in the
right-hand part of Fig.~\ref{fig:Hubbard}.
The lower group conserves the structure of levels (a non-degenerate
ground state and doubly-degenerate first excited state), 
although the energy gap is bigger $\Delta E^{Hub} = 0.0111 \ Ry^*$. 
On the other hand the upper group of levels not only shows bigger
energy gaps, but the fifth and sixth excited states, nondegenerate in
the full LCHO-CI calculation, become degenerate.

\subsection{Triple dot under bias} \label{bias} 

\begin{figure}[h]  
\includegraphics[width=0.8\textwidth]{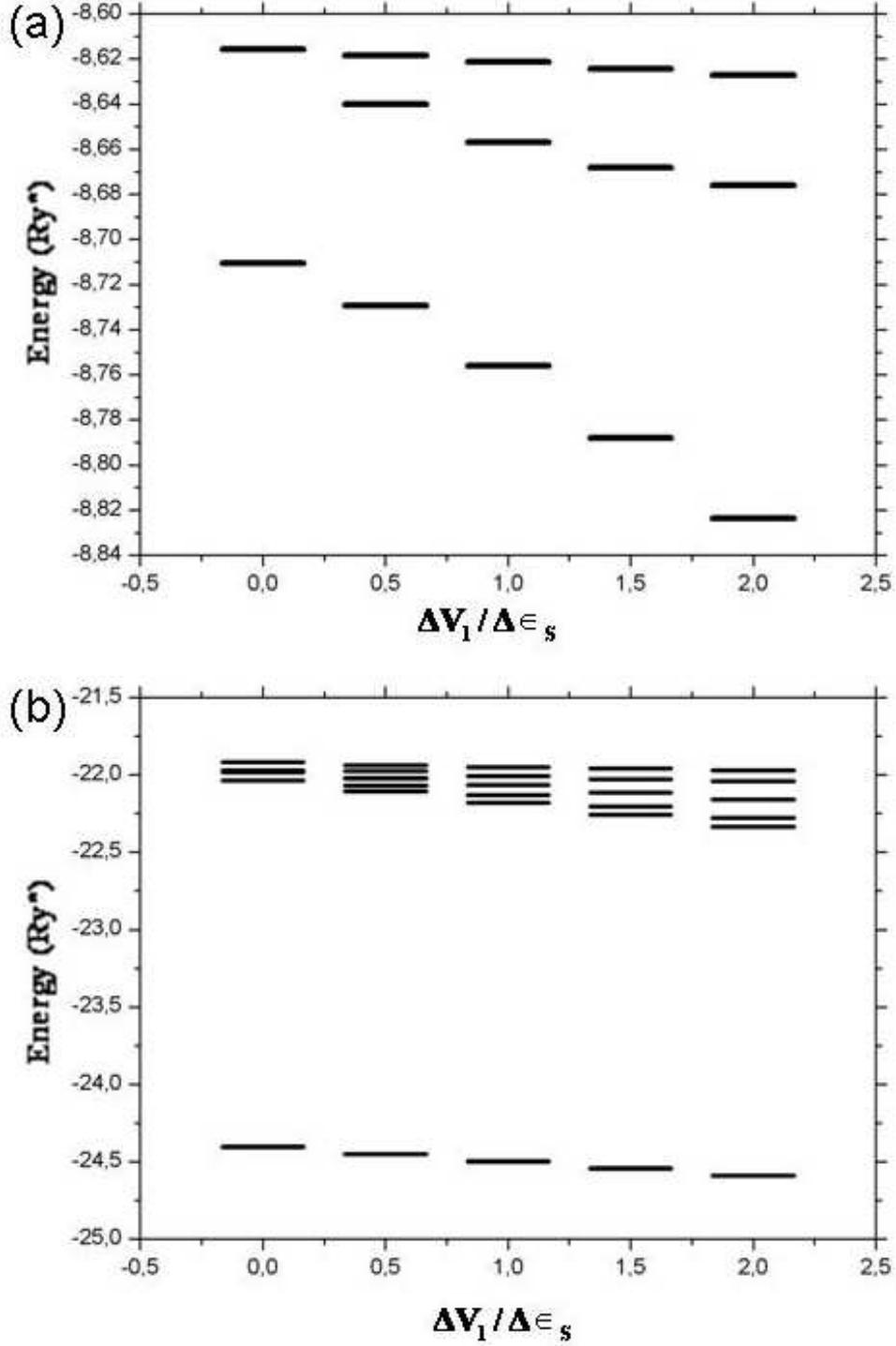}
\caption{(a) Spectra of one electron in the triple dot as a function
  of bias of dot 1.   
  (b) Spectra of three electrons with $S_z =-1/2$ in the triple dot a
  function of bias of dot 1, calculated with the LCHO-CI method
  considering $s$-HO levels in the LCHO calculation.}  
\label{fig:1elEnBiasDot1} 
\end{figure}

In this section we study the evolution of the energy spectrum of a
three-electron molecule as a function of $V_1$, i.e., the depth of dot
$1$. 
The applied  bias  is kept  smaller than the $s$-$p$ energy gap of one
electron in one quantum dot in order to prevent population of the
biased dot with two electrons.  
This is done so that we can attempt to map the three electron spectra
onto the three spin  spectra obtained in the Heisenberg model. 
We carry out our calculations for the case of $s$-HO orbitals as LCHO
basis.  
Figure \ref{fig:1elEnBiasDot1} (a) shows the resulting one-electron 
spectra as a function of bias of dot 1. 
The bias is measured in the units of the one-electron energy gap  
$\Delta \epsilon_S = 0.095$~$Ry^*$.  
When the dot is biased, the degenerate excited levels split due to the
breaking of symmetry.  
This energy splitting increases with increasing bias .  
It is also observed that all energies decrease, but each of the three
energies at a different rate.  
This can be understood by analyzing the evolution of orbitals
associated with these energies.
When the three dots are on resonance, the two excited levels are  
degenerate, and any linear combination of the two eigenfunctions 
$|\xi_2 \rangle$ and  $| \xi_3 \rangle$, defined in
Eq.~(\ref{eq:1elef}), is also an eigenfunction of the
system corresponding to the same energy.  
This leads to equal probability of finding of the electron in each of
the dots.
However, if one of the dots is different, the excited level splits into  
two, and the wave functions $| \xi_2 \rangle$ and $| \xi_3 \rangle $
reflect this symmetry breaking. 
The first excited state $| \xi_2 \rangle $ has a contribution from the
HO orbital corresponding to the first dot that decreases   
with increasing bias $V_1$ (at the same time the contribution of this  
orbital increases in the ground state). 
On the other hand, the second excited state $| \xi_3 \rangle $ does
not have any contribution from this orbital.    

Figure \ref{fig:1elEnBiasDot1}(b) shows the spectra of three electrons
with $S_z =-1/2$ in the triple dot molecule for different values of
$V_1$.  
We are interested in the three lowest levels as these are the ones  
relevant for quantum information processing and thus the ones that
will be mapped onto Heisenberg spectra.  
Figure \ref{fig:Zoom3elBiasV1Heis}(a) shows the energies of these
levels measured from the ground state energy.
The degenerate ground state splits with bias. 
When bias is increased, the energy gap between the ground state and
the first excited state increases, while the energy gap between the
first and second excited states decreases.    

\subsection{Heisenberg model for three electrons} \label{heis} 

We now attempt to map the three-electron spectrum for the case in
which dot 1 is biased onto the spectrum of the three-spin 
Heisenberg Hamiltonian (\ref{eq:1/2-2}) in the Jacobi basis. 
The mapping relies on the assumption that biasing dot $1$ can be
modeled by bias-dependent but equal parameters $J_{12}=J_{13}=J$,
while $J_{23}$ is different, but it may also be a function of bias. 
With $J_{12}$ equal $ J_{13}$, the two Jacobi states $|\beta_a\rangle$
and $|\beta_b\rangle$ are eigenstates, with energies  
$-\frac{3}{4} J_{av} -\frac{3}{4} ( J_{23}-J_{av} )$ 
and $-\frac{3}{4} J_{av} +\frac{3}{4} ( J_{23}-J_{av} )$, 
respectively, and $|\beta_c\rangle$ is an eigenstate with energy  $+
\frac{3}{4}J_{av}$. 
The three eigenstates can be written in a way which emphasizes the
role of the electron in the dot under bias (dot 1):  
$|\beta_a \rangle  =   |\downarrow  \rangle| S  \rangle  $ and 
$|\beta_b \rangle = 
1/ \sqrt{3}   | \downarrow  \rangle| T_0  \rangle 
-\sqrt{2/3}   | \uparrow  \rangle| T_-  \rangle $.  
Here  the singlet and triplet states involving the second and third
dot were written as 
$| S  \rangle = 1/\sqrt{2} 
(|\downarrow \uparrow \rangle - |\uparrow \downarrow \rangle)$,  
$| T_0  \rangle = 1/ \sqrt{2} (|\downarrow \uparrow \rangle 
+ |\uparrow \downarrow \rangle) $ and 
$| T_-  \rangle = |\downarrow \downarrow \rangle $. 

\begin{figure}[h]  
\includegraphics[width=0.8\textwidth]{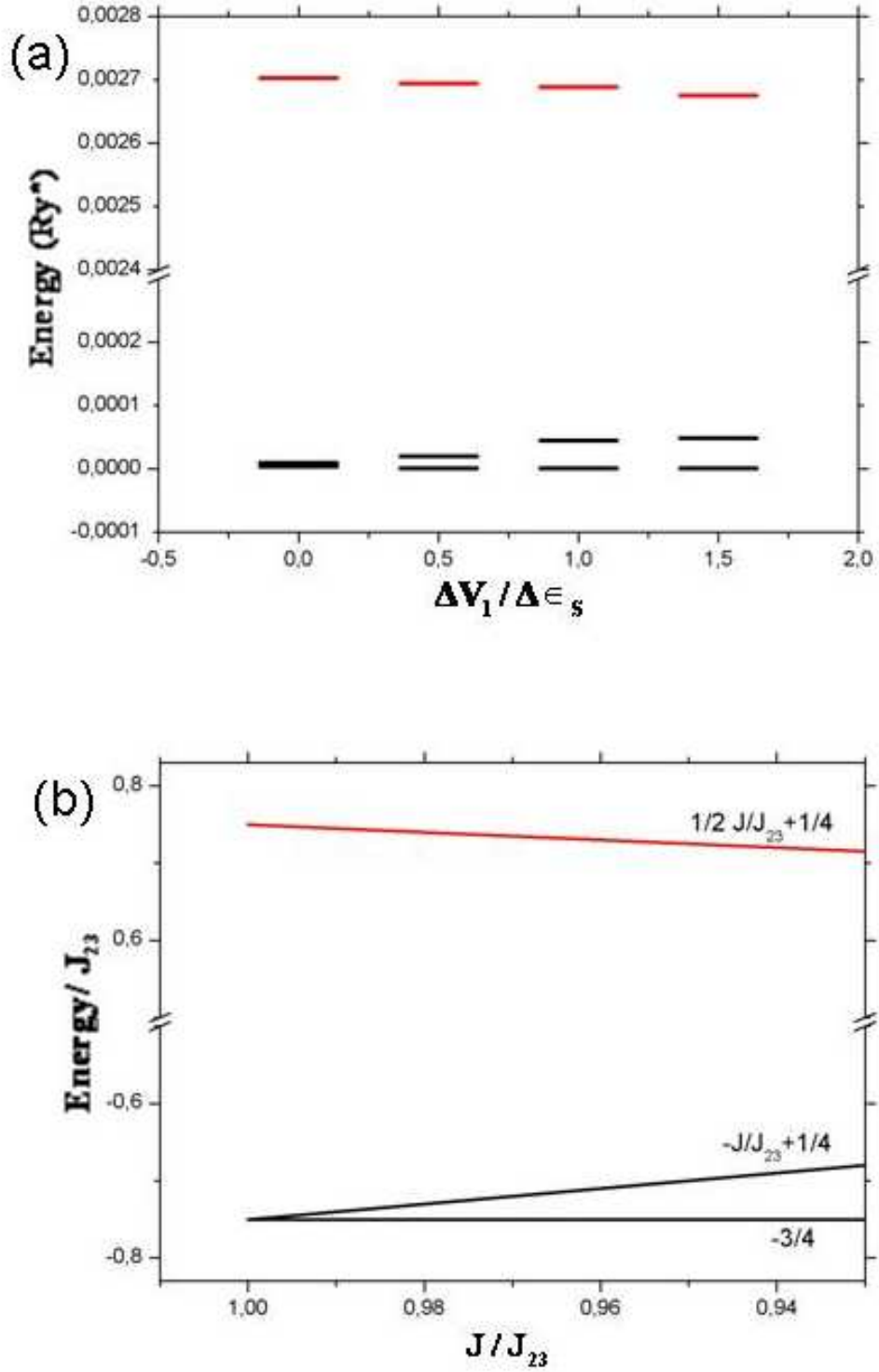}
\caption{(a) Three lowest levels of spectra of three electrons in
  the triple dot (Fig. \ref{fig:1elEnBiasDot1}), with the ground state
  energy as a reference level, as a function of bias of dot 1.
  (b) Spectra of three electrons $S_z =-1/2$ given by
  Heisenberg model with $J_{12}=J_{13} \equiv J$ plotted 
  as a function of $J/J_{23}$ with $J$ varying from $J_{23}$ to $0.93
  J_{23}$. }  
\label{fig:Zoom3elBiasV1Heis} 
\end{figure}

Figure \ref{fig:Zoom3elBiasV1Heis}(b) shows a plot of the energy
spectrum of  the Heisenberg Hamiltonian 
as a function of $J/J_{23}$ from $J=J_{23}$ to $J=0.93 J_{23}$ with
$J>0$ (antiferromagnetic ground state). 
By changing $J$ from $J_{23}$ to $J=0$ we drive the system from three
equal dots to dot 1 totally decoupled. 
We compare this spectrum to that obtained in the full electronic
calculation, shown in Fig.~\ref{fig:Zoom3elBiasV1Heis}(a).
As we can see, the two spectra behave in a similar manner as a
function of bias of one of the dots ranging from three   
equal dots (zero bias) to a bias of $\sim 1.5$ times the one-electron energy
gap.
We can now propose a procedure of finding the parameters $J_{ij}$ by
fitting the energy gaps in the Heisenberg spectrum,
$\Delta \varepsilon ^{s}_{ab}=J_{23}-J$ and 
$\Delta \varepsilon ^{s}_{bc}=\frac{3}{2}J$, to those in the
LCHO-CI calculation.
The value of $J_{23}$ can be extracted at zero bias from the gap
between the doubly-degenerate ground and the excited states.
Assuming $J_{23}$ to be independent of the bias of dot 1, 
two values of $J$ - one from each gap of the spectrum -
can be obtained and averaged for each step of bias.  
The numerical results show a decreasing value of the coupling constant
$J$ as the bias is being increased.  
This is the expected behavior because as we bias dot 1 the exchange
of the electron in  dot 1 with electrons in dots 2 and 3  decreases.  

\section{Conclusions} 

We presented a computational LCHO-CI approach allowing for the 
simulation of exchange interactions in  gated lateral  quantum 
dot networks. 
The method was illustrated by analyzing the electronic properties of a
lateral triple quantum dot network with one electron per dot. 
The LCHO-CI calculations show a low-energy spectrum composed of an
antiferromagnetic ($S=1/2$) ground state separated by a small gap from 
the spin polarized ($S=3/2$)  excited state, and separated by a large
gap from the remaining excited levels involving double occupancy  
of  quantum dots. 
We have shown that the behavior of these eight low-energy levels with
bias of one quantum dot can be effectively reproduced by a Heisenberg
spin model for a certain range of bias applied to quantum dots. 
For each value of bias, exchange coupling constants can be calculated
from energy gaps in LCHO-CI model. 
We have thus established a connection between physical ''external''
parameters such as  voltages, and exchange interaction among spins in
the Heisenberg model.

\section{Acknowledgment}

We wish to thank A. Sachrajda, L. Gaudreau, A. Kam, and S. Studenikin
for helpful discussions.
The authors are grateful to the Canadian Institute for Advanced Research,
and IPG to Spanish Ministerio de Educaci\'on y Ciencia grant
No. AP-2004-0143 for financial support.

\newpage



\end{document}